\documentclass[a4paper, 11pt]{article}
\usepackage{pos}

\title{A Search for Neutrino Sources with Cascade Events in IceCube}

\ShortTitle{A Search for Neutrino Sources with Cascade Events in IceCube}

\author{The IceCube Collaboration \\
{\normalsize \normalfont(a complete list of authors can be found at the end of the proceedings)}}

\emailAdd{ssclafani@icecube.wisc.edu}
\emailAdd{mhuennefeld@icecube.wisc.edu}

\abstract{

IceCube has discovered a flux of astrophysical neutrinos, and more recently has used muon-neutrino datasets to present evidence for one source; a flaring blazar known as TXS 0506+056. However, the sources responsible for the majority of the astrophysical neutrino flux remain elusive. Opening up new channels for detection can improve sensitivity and increase the discovery potential. In this work we present a new neutrino dataset relying heavily on Deep-Neural-Networks (DNN) to select cascade events produced from neutral-current interactions of all flavors and charged-current interactions with flavors other than muon-neutrino. The speed of DNN processing makes it possible to select events in near realtime with a single GPU. Cascade events have reduced angular resolution when compared to muon-neutrino events, however the resulting dataset has a lower energy threshold in the southern sky and a lower background rate. These benefits lead to an factor of 2-3 improvement in  sensitivity to sources in the Southern Sky when compared to muon-neutrino datasets. This dataset is particularly promising for identifying transient neutrino sources in the Southern Sky and neutrino production from the galactic plane.\\

\vspace{4mm}
{\bfseries Corresponding authors:}
{Stephen Sclafani}$^{1*}$, Mirco H{\"u}nnefeld$^{2}$\\
{$^{1}$ \itshape Drexel University}\\
{$^{2}$ \itshape Technische Universit{\"a}t Dortmund}\\[4mm]
{$^*$ Presenter}
}

\FullConference{37$^{\rm{th}}$ International Cosmic Ray Conference (ICRC 2021)\\
		July 12th -- 23rd, 2021\\
		Online -- Berlin, Germany}

\begin{document}
\maketitle

\section{Introduction}\label{sec:intro}
In 2017, the IceCube Neutrino Observatory, a cubic kilometer detector located at the geographic south pole, hinted at the long-sought sources of Cosmic Rays by presenting evidence of the first neutrino source, the flaring blazar TXS0506+056 \cite{IceCube:2018cha, IceCube:2018dnn}.  This detection was done by analyzing data made up of events produced from Charge-Current muon-neutrino interactions creating a long ranged muon that leaves a track-like signature in the detector. TXS0506+056 is located in the Northern Sky, at a declination of +5.6 degrees which is in the optimal region for the search of point sources using muon neutrinos. The Southern Sky, however, is dominated by the large contribution of atmospheric muon background strongly reducing the discovery potential for sources with soft spectral indices ($\gamma$ $>$ 2). There exists two strategies to advance the detection of the next neutrino source: accumulating more statics will slowly but steadily improve the sensitivity for time-integrated searches, additionally, a new detection channel could help to accelerate the discovery.  Including cascade topologies in the event selection will extend the search for point sources in the Southern Sky while complementing the results of the muon track searches. Cascades are produced from neutral current interactions from all neutrino flavors and charge current interactions from  flavors other than muon neutrinos.  These events have worse angular resolution, but can be easily differentiated from the dominant background of downing muons from atmospheric interactions in the Southern Sky.  Thus, selections based on cascade events, will have a higher level of purity even at lower energies (500\,GeV $<$ E $< $ 10\,TeV).  This results in an improvement in sensitivity to southern sources, especially sources with softer spectra, which has been demonstrated in a previous IceCube search \cite{Aartsen:2019epb} using a cascade dataset designed for diffuse measurements.  This work uses machine learning to create a new cascade dataset optimizing selection to improve sensitivity to several different neutrino source hypotheses.

\section{Data Selection}\label{sec:dataset}
The event selection pipeline consists of a series of DNN classifiers that are based on previous IceCube work \cite{Abbasi:DNNRECO}. These classifiers are applied to select events starting at "Cascade Level 2", a very basic selection without any advanced reconstruction or filtering. Although these classifiers are very quick to evaluate, the event rate for IceCube at Level 2 is very large (30-40 Hz) and therefore it is important to reduce the event rate and discard background events before applying more time-intensive reconstructions.  For this reason, events are classified in a staged approach, with more advanced and time consuming algorithms running only on events that passed the previous step.  During processing, a series of NN classifiers are applied to incrementally reduce the background while increasing cascade purity.   

\begin{figure}
    \centering
    \includegraphics[width=0.5\textwidth]{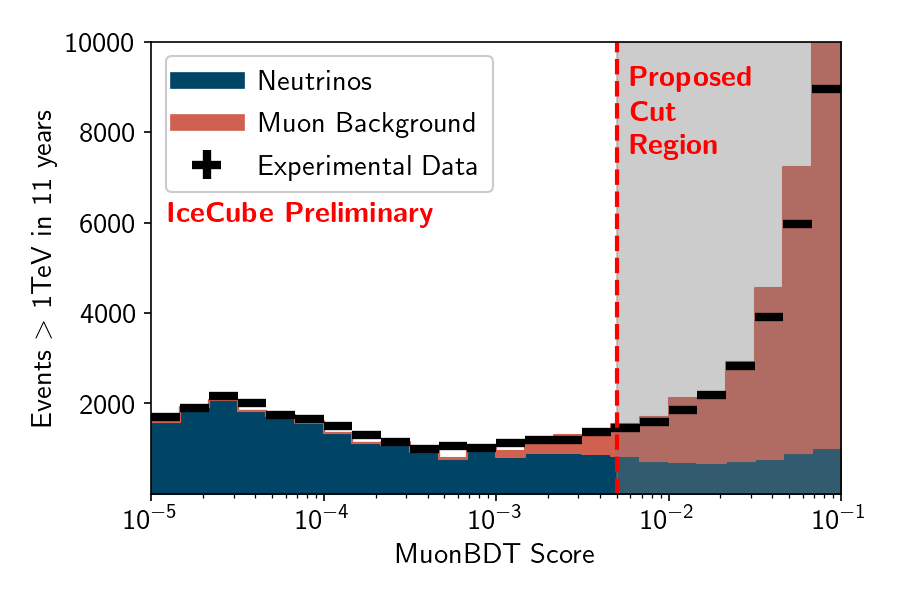}
    \caption{The score distribution of events with reconstructed energy greater than 1\,TeV for the BDT trained to classify background muons. Proposed cut is currently placed at 5 $\times$ 10$^{-3}$. To the right of this value, selection is dominated by muon background and Data/Monte Carlo agreement worsens. The exact location of this cut is still under investigation.}
    \label{fig:muonBDT}
\end{figure}

\begin{figure}
    \centering
    \includegraphics[width=0.5\textwidth]{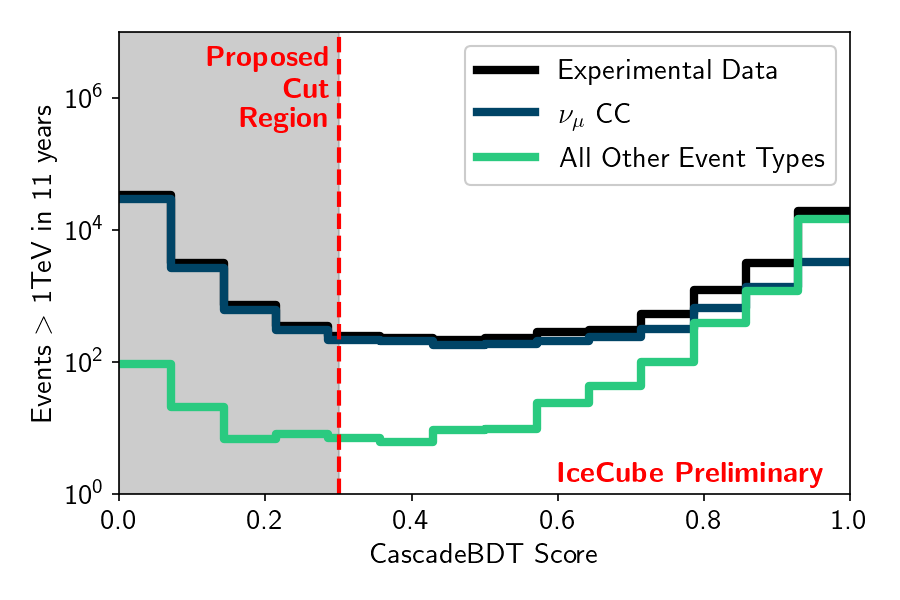}
    \caption{The score distribution of events with reconstructed energy greater than 1\,TeV for the BDT trained to classify cascades. Proposed cut is currently placed at 0.3. To the left of this value, selection is dominated by tracks produced from charge-current muon-neutrino interactions. To the right cascades from other event types become dominant, however CC-$\nu_\mu$ events that are particularly stochastic, and whose track length is short, still remain in the sample. The exact location of this cut is still under investigation.}
    \label{fig:cascadeBDT}
\end{figure}

The final selection is made using two Boosted Decision Trees (BDTs).  The first (muonBDT) is trained to remove any remaining atmospheric muon background (Figure \ref{fig:muonBDT}).  The second (cascadeBDT) is trained to select cascade-like events (Figure \ref{fig:cascadeBDT}). Proposed cuts are placed on each BDT separately, and can be combined with cuts on energy or angular error estimate. Final cut values will be optimized based on discovery potential and sensitivity at analysis level and are still under investigation. As can be see in Figure \ref{fig:muonBDT}, events with a high muonBDT score are dominated by background events. We also see that the agreement between data and simulation worsen in this background dominated regime.  Some charge-current muon-neutrino events with very short and stochastic tracks, or tracks that exit the detector quickly may remain in the sample after cuts. The total selection takes a few seconds per event.  The result is a dataset made up of approximately 45,000 cascade-like events from 2011 -- 2021.  We estimate that the muon background will be $\mathcal{O}(100)$ events at final level. 
    
After the final cut, a time-intensive reconstruction (30s) can take place. The reconstruction combines a Maximum Likelihood Method and DNN based reconstructions \cite{EventGenerator}, resulting in a improvement in the angular resolution, especially in events less that 10\,TeV.  

\begin{figure}
    \centering
    \includegraphics[width=0.5\textwidth]{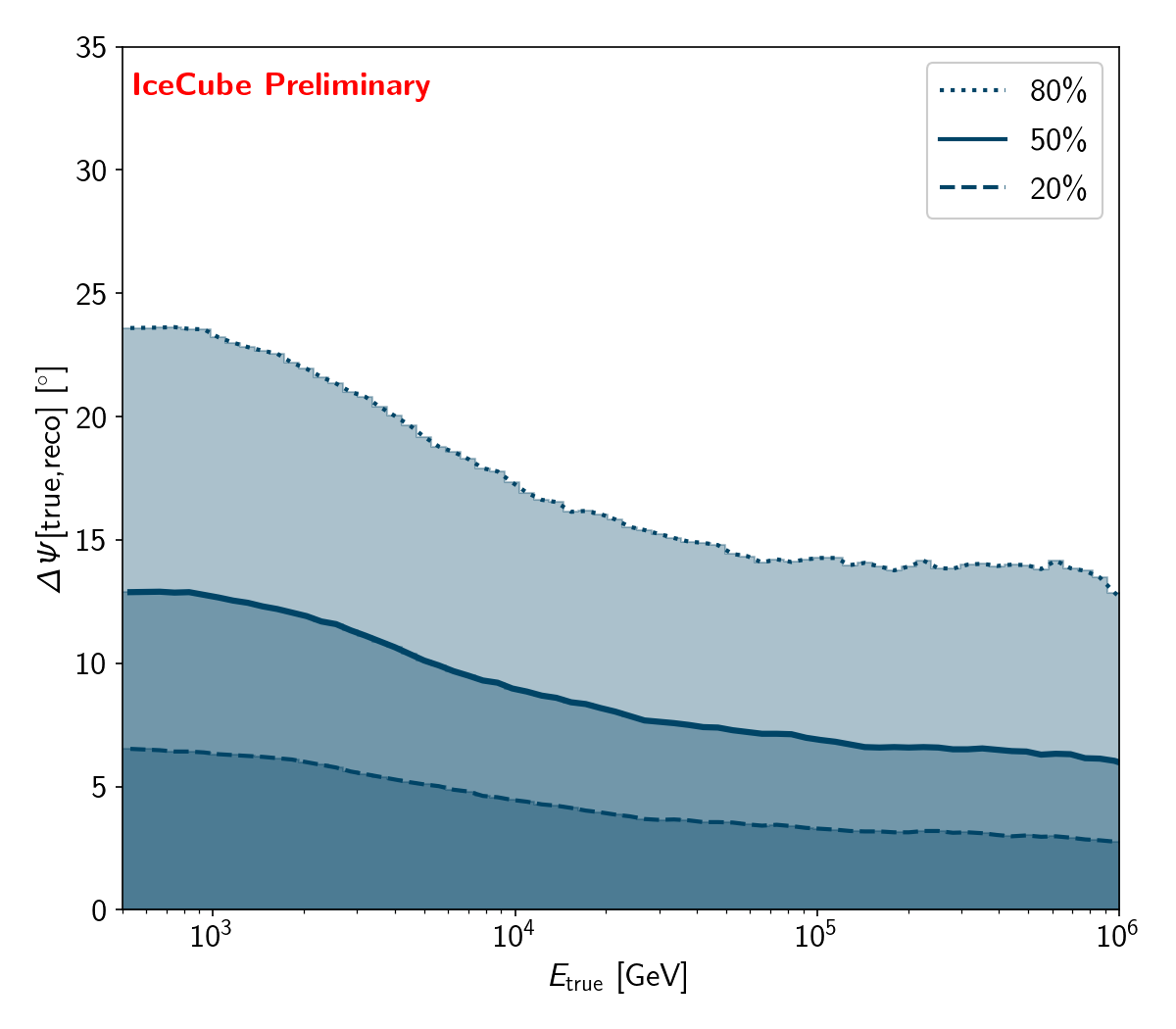}
    \caption{Angular Resolution as a function of energy. Events get better resolved as energy increases.}
    \label{fig:angres}
\end{figure}

\begin{figure}
    \centering
    \includegraphics[width=0.5\textwidth]{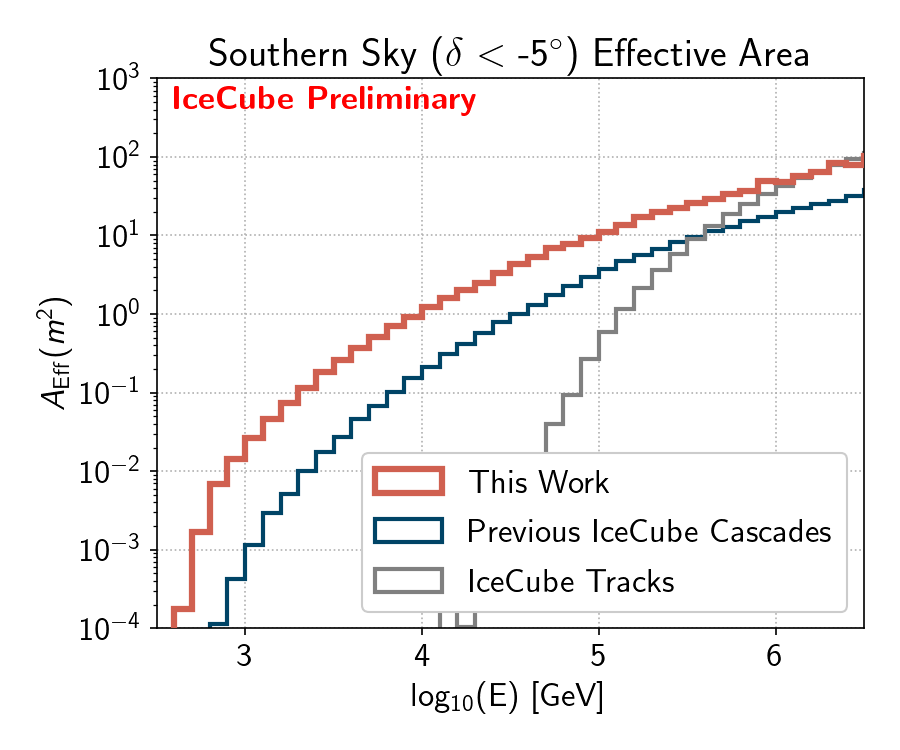}
    \caption{Effective area of this dataset compared with IceCube tracks and previous cascade dataset for declinations in the southern sky.}
    \label{fig:effa}
\end{figure}

 Figure \ref{fig:angres} shows the final level angular resolution as a function of true neutrino energy.  The median angular resolution is between 5$^\circ$ and 15$^\circ$ over the range of energies in the dataset. The resulting dataset has a larger effective area over all energies than previous cascade datasets, and a larger effective area in the Southern Sky than the track-based dataset \cite{Aartsen:2019fau} shown in Figure \ref{fig:effa}.   

\section{Point Source Searches}\label{sec:ps_searches}
The event selection described above, named DNNCascade, will be used to perform various searches for the sources of astrophysical neutrinos.  Because of the low atmospheric muon background even at low energies, these searches are particularly promising for sources in the Southern Sky, sources with a softer spectrum such as galactic sources, or sources an exponential cutoff in the TeV-PeV range.  

\begin{figure}
\includegraphics[width=.33\textwidth]{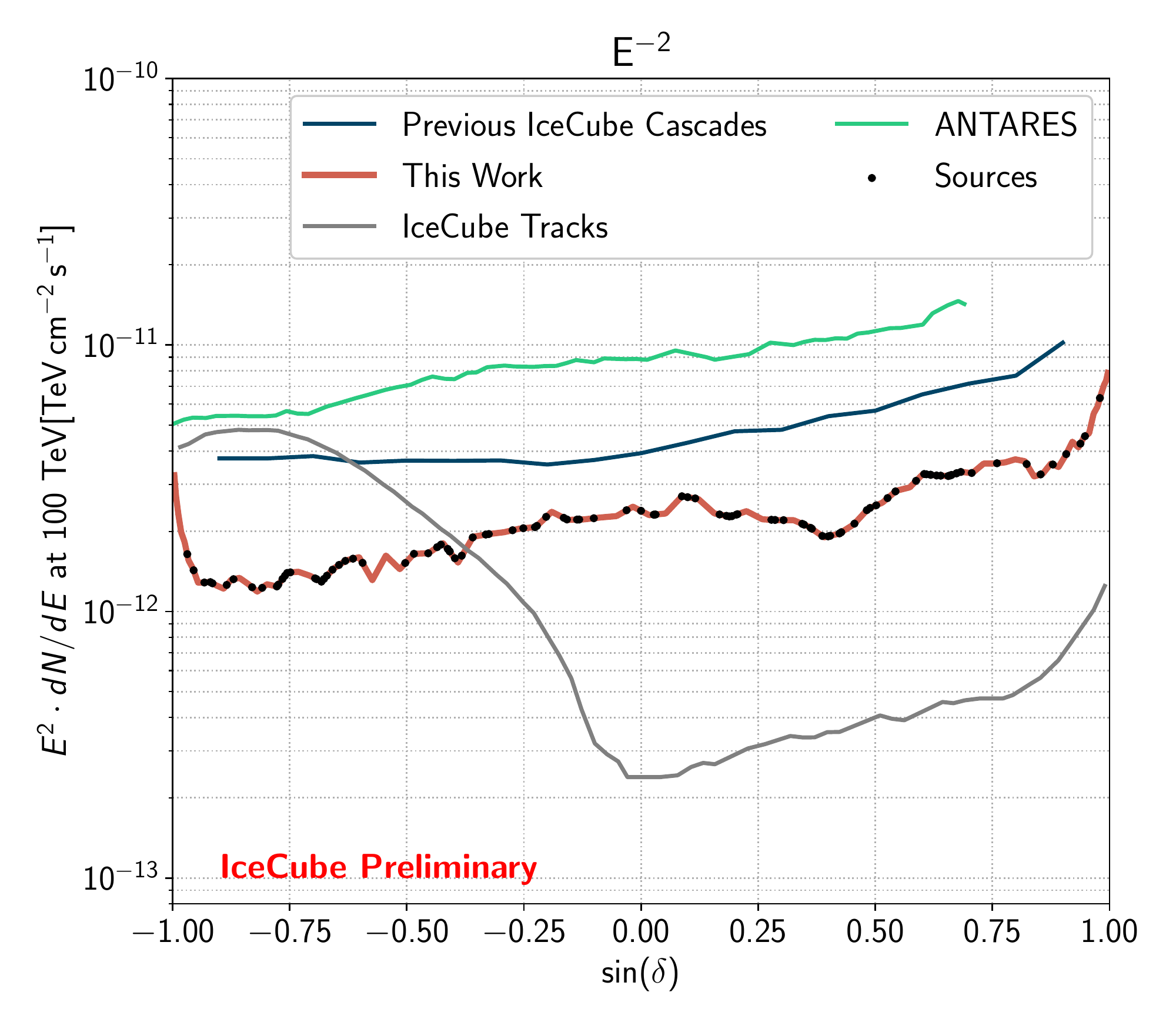}
\includegraphics[width=.33\textwidth]{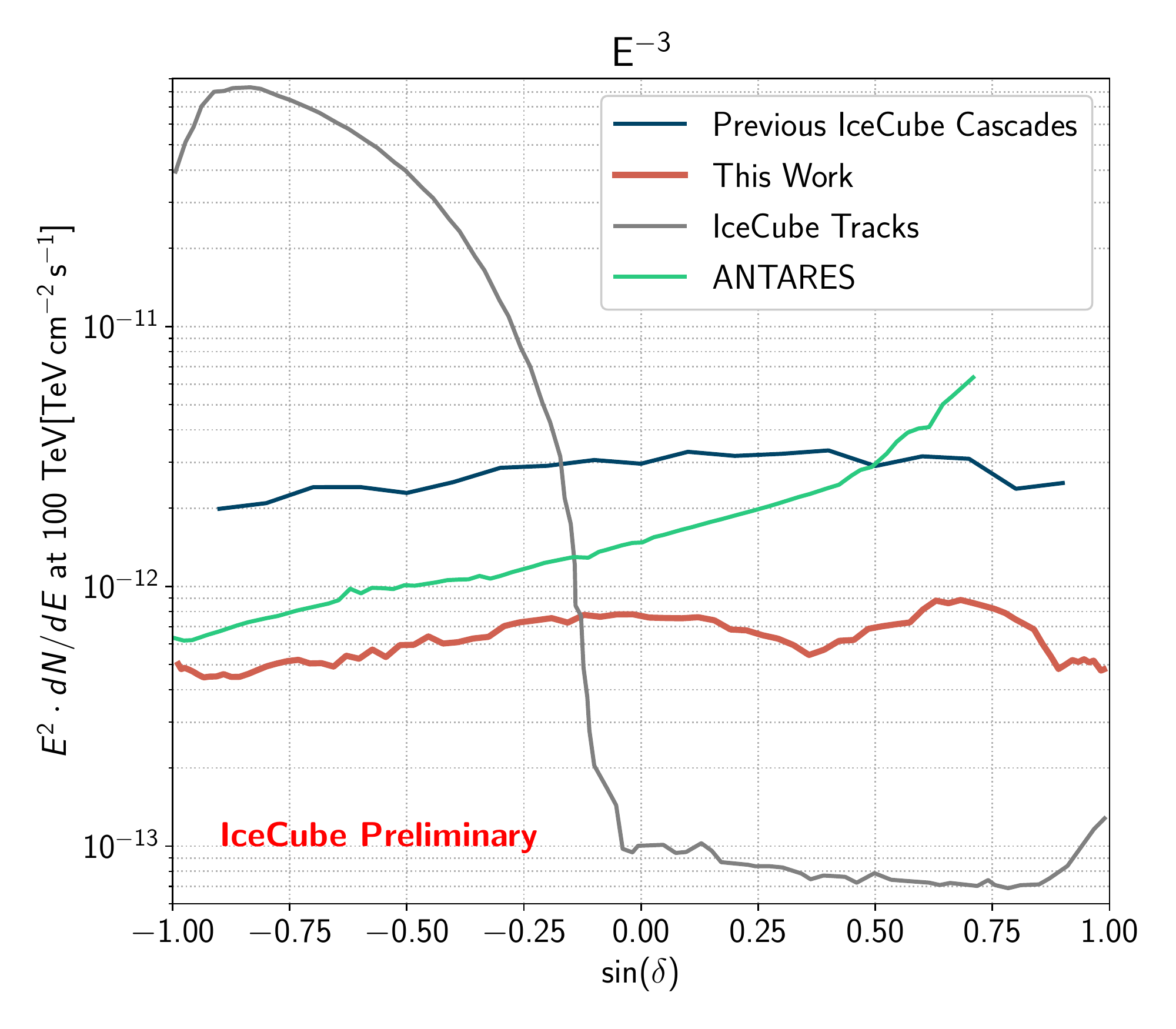}
\includegraphics[width=.33\textwidth]{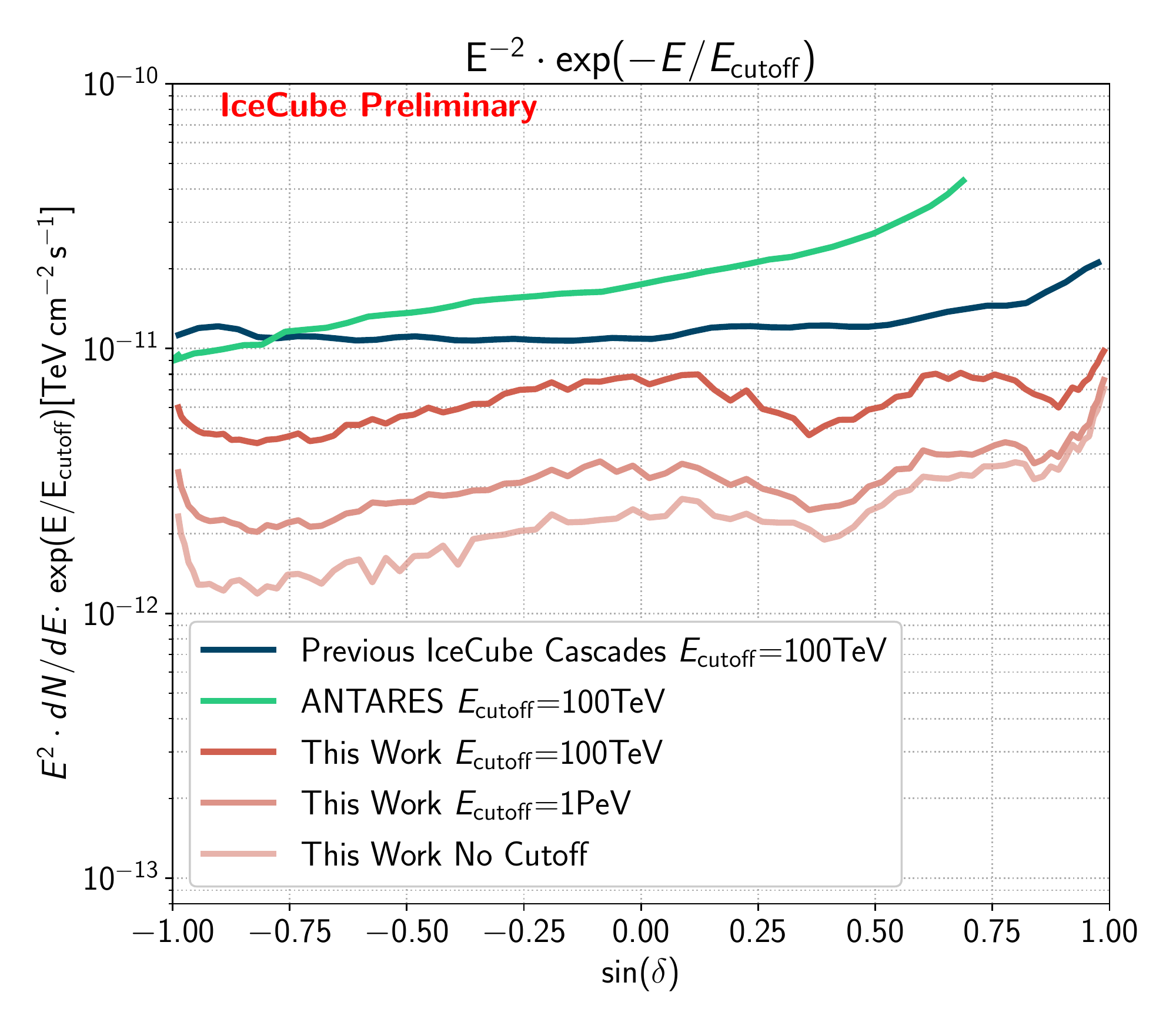}

\caption{Per-flavor sensitivity as a function of sin($\delta$) to point sources following an unbroken E$^{-2}$ spectrum (left), unbroken E$^{-3}$ spectrum (center), and E$^{-2}$ spectrum with some possible exponential cutoffs (right).  Comparisons are made to ANTARES \cite{Illuminati:2019i3} for E$^{-2}$ and E$^{-3}$  and \cite{Albert:2017ohr} for the exponential cutoffs, IceCube Tracks \cite{Aartsen:2019fau}, and previous IceCube cascades  \cite{Aartsen:2019epb}. Declination of sources to be tested are shown as black dots on the left plot.}
\label{fig:ps_sens}
\end{figure}

Figure \ref{fig:ps_sens} shows the sensitivity flux as a function of sin of declination for a source spectrum proportional to E$^{-2}$, 
E$^{-3}$, and E$^{-2}$ with a particular exponential cutoff.  Comparisons are made to previous IceCube \cite{Aartsen:2019epb, Aartsen:2019fau} and ANTARES \cite{Illuminati:2019i3, Albert:2017ohr} sensitivities.  Sensitivities to specific sources from a catalog that will be used in the future analysis are also shown in Figure \ref{fig:ps_sens} (left panel). This catalog of sources is based on the expected neutrino flux inferred from  $\gamma$-ray observations and the declination dependent sensitivity of this dataset.  

\section{Stacking Searches}

Stacked searches consider the total flux emitted by a number of sources from the same source class.  This analysis strategy has the advantage that, if all sources from the same class are neutrino emitters, the neutrino flux per source required for discovery is lower.  Stacking searches have been used in IceCube and other neutrino experiments to search for emission from catalogs of similar classes of sources.  Three catalogs of galactic $\gamma$-ray emitters are tested consisting of 12 Supernova Remnants (SNR) , Pulsar Wind Nebulae (PWN), and Unidentified Galactic objects (UNID).  The sources are selected based on an extrapolation of the $\gamma$-ray flux above 10\,TeV and the sensitivity to this dataset.  For this test all sources are given equal weight before any detector effect is taken into account.  Table \ref{table:stacking} shows the sensitivity to these source catalogs assuming different spectra. The catalogs are available in Table \ref{table:source_list_stacking}.  In these searches we will fit for the number of signal events and energy spectrum.  

\begin{table}[!ht]
\centering
\begin{tabular}{|c|c|c|c|}
\hline
Source Catalog & Sensitivity E$^{-2}$ & Sensitivity E$^{-2.5}$  \\
\hline 
SNR &  3.39 & 3.31  \\
PWN & 3.99 & 3.51  \\
UNID Galactic Sources &  3.85  & 3.18\\
\hline
\end{tabular}
\caption{Per-Flavor Stacked Sensitivity to various catalogs of Galactic Sources.  Sensitivity shown as E$^2$ dN/dE $\times$ 10$^{-12}$ at 100\,TeV in units of TeV $\cdot$  s$^{-1}$ $\cdot$ cm$^{-2}$.}
\label{table:stacking}
\end{table}

\section{Galactic Plane Searches}\label{gp_searches}

\begin{figure}
    \centering
    \includegraphics[width=0.5\textwidth]{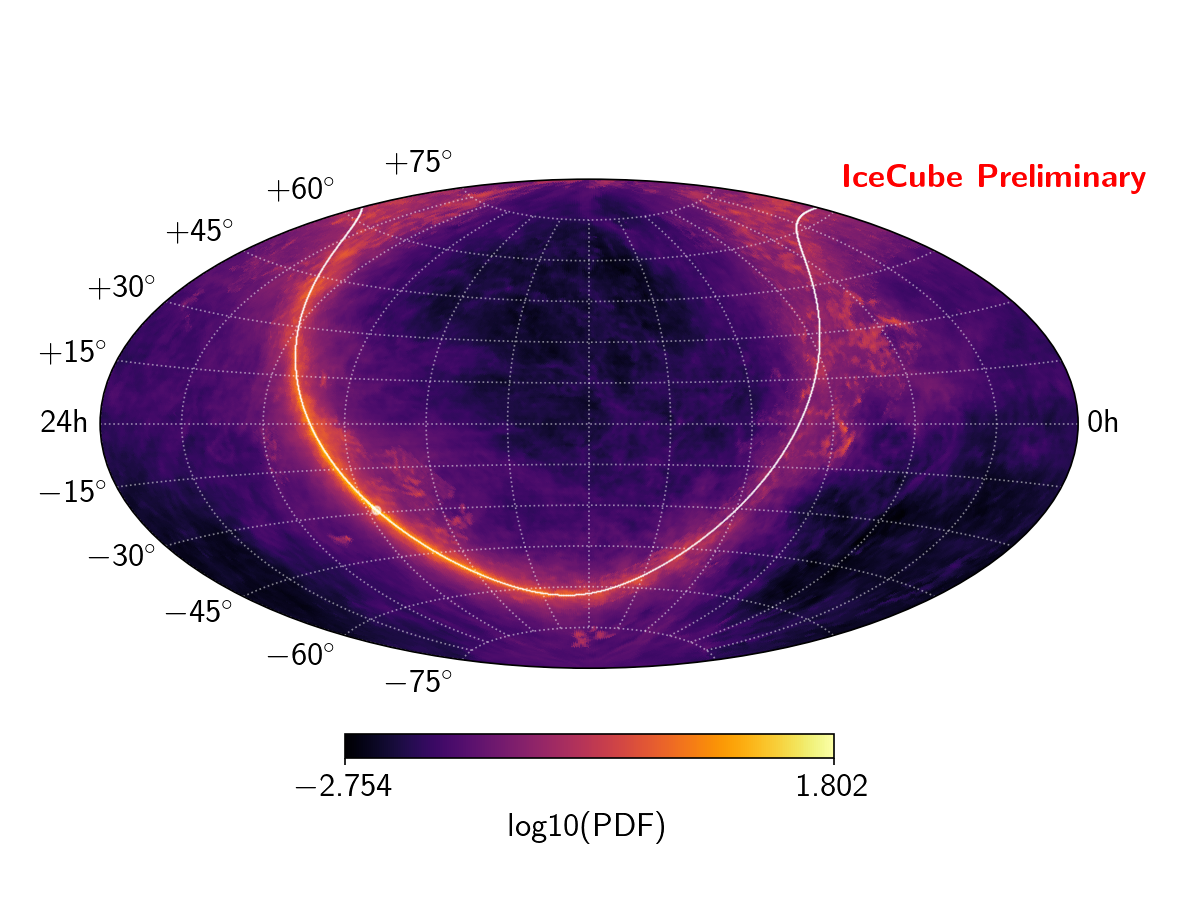}
    \caption{The spatial template for the KRA-$\gamma$ model with 5\,PeV cutoff. The model is integrated over all energies and multiplied with the declination and spectra dependent signal weighting parameter for this dataset. The resulting PDF does not include any additional spatial smoothing. Shown in white is the galactic plane, the white dot is the galactic center. Expected Flux is concentrated along the plane, closest to the galactic center.}
    \label{fig:gp}
\end{figure}

Cosmic-ray interactions with galactic media is  the dominant production mechanism for high energy gamma rays through the production of neutral pions.  A flux of neutrinos from galactic plane is expected from these same interactions through the production of charged pions. Models predict that the neutrino flux would be concentrated in the Southern Sky and roughly follow an E$^{-2.5}$ power law. The DNNCascade dataset is optimized to find sources with the same properties and tests for diffuse galactic emission will be performed. Three models are tested, a model based on Fermi-LAT observations \cite{Ackermann_2012} and KRA-$\gamma$ \cite{Gaggero:2015xza, PhysRevLett.119.031101} with a cutoff of 5\,PeV and KRA-$\gamma$ with a 50\,PeV cutoff.  These rely on $\gamma$-ray information to construct a template for the expected neutrino production.  The KRA-$\gamma$ templates have a direction dependant spectrum and is shown in Figure \ref{fig:gp}.  This template contains most of the weight at near the galactic center in the Southern Sky.\\

Sensitivities to each model are presented in Table \ref{table:gp_sens}.  Comparisons to previous results of IceCube using cascades \cite{Aartsen:2019epb} and combined analysis of IceCube and ANTARES \cite{Albert:2018vxw} using IceCube tracks are also shown.  Sensitivities of this work do not include effects of systematic uncertainties. For all models of galactic emission this work is the most sensitive and represents a significant improvement in our ability to probe the presented models.  In addition, the 5$\sigma$ discovery potentials are calculated to be 69\% and 49\% of the model flux for KRA$\gamma$- with a 5\,PeV and 50\,PeV cutoff respectively.  \\

\begin{table}[!ht]
\centering
\begin{tabular}{|c|c|c|c|}
\hline
Galactic Plane Model & \textbf{This Work} & IceCube / ANTARES & Previous IceCube Cascades \\
\hline 
KRA-$\gamma$ 5\,PeV & 0.17 & 0.81 & 0.58 \\
KRA-$\gamma$ 50\,PeV  & 0.12 & 0.57 & 0.35 \\
Fermi $\pi^{0}$ &  0.82 $\times$ 10$^{-18}$ & -- & $2.2 \times 10^{-18}$ \\
\hline
\end{tabular}
\caption{Sensitivity to various galactic plane models.  KRA-$\gamma$ templates are in units of model flux.  Fermi $\pi^0$ are in GeV s$^{-1}$cm$^{-2}$ at 100\,TeV. IceCube/ANTARES sensitivity is from \cite{Albert:2018vxw}, Previous IceCube Cascade Sensitivity is from \cite{Aartsen:2019epb}.  This work does not include effects of systematic uncertainties. }
\label{table:gp_sens}
\end{table}

\section{Conclusions}

IceCube is extending its reach by considering cascade-like events in the search for neutrino sources.  This new channel improves the sensitivity in the Southern Sky by a factor of 3 for soft spectrum sources ($\gamma$ $=$ 3) and galactic plane emission, by a factor of 2 for sources with a harder spectral index ($\gamma$ $=$ 2). This dataset will be used in the future for time-dependant searches, where significant detections can occur without precise pointing. The speed of the dataset processing due to the DNN structure will make it possible to select events online with few resources, opening the window to real-time analyses using these cascades.  

\bibliographystyle{ICRC}
\bibliography{references}
\appendix
\newpage
\section{Source Catalogs}
\begin{table}[!htb]
\centering
\small
\begin{tabular}{|c c c c| c c c c |}
\hline
Catalog & Name & $\alpha$[deg] & $\delta$[deg]  & Catalog & Name & $\alpha$[deg] & $\delta$[deg] \\
\hline 
PWN & Vela X & 128.29 & -45.19 & SNR & Vela Junior  & 133.0 & -46.33 \\ 
 & Crab nebula & 83.63 & 22.01 & & RX J1713.7-2946 & 258.36 & -39.77 \\ 
 & HESS J1708-443 & 257.0 & -44.3 & & HESS J1614-518 & 243.56 & -51.82 \\ 
 & HESS J1825-137 & 276.55 & -13.58 & & HESS J1457-593 & 223.7 & -59.07 \\ 
 & HESS J1632-478 & 248.01 & -47.87 & & SNR G323.7-01.0 & 233.63 & -57.2 \\ 
 & MSH 15-52 & 228.53 & -59.16 & & HESS J1731-347 & 262.98 & -34.71 \\ 
 & HESS J1813-178 & 273.36 & -17.86 & & Gamma Cygni & 305.27 & 40.52 \\ 
 & HESS J1303-631 & 195.75 & -63.2 & & RCW 86 & 220.12 & -62.65 \\ 
 & HESS J1616-508 & 244.06 & -50.91& & HESS J1912+101 & 288.33 & 10.19 \\ 
 & Kookaburra & 214.69 & -60.98 & & HESS J1745-303 & 266.3 & -30.2 \\ 
 & HESS J1837-069 & 279.43 & -6.93 & & Cassiopeia A & 350.85 & 58.81 \\ 
 & HESS J1026-582 & 157.17 & -58.29 & & CTB 37A & 258.64 & -38.545 \\ 
 \hline
 UNID & MGRO J1908+06 & 133.0 & 6.32 \\ 
 & Westerlund 1 & 258.36 & -45.8 \\ 
 & HESS J1702-420 & 243.56 & -42.02 \\ 
 & 2HWC J1814-173 & 223.7 & -17.31 \\ 
 & HESS J1841-055 & 233.63 & -5.55 \\ 
 & 2HWC J1819-150 & 262.98 & -15.06 \\ 
 & HESS J1804-216 & 305.27 & -21.73 \\ 
 & HESS J1809-193 & 220.12 & -19.3 \\ 
 & HESS J1843-033 & 288.33 & -3.3 \\ 
 & TeV J2032+4130 & 266.3 & 41.51 \\ 
 & HESS J1708-410 & 350.85 & -41.09 \\ 
 & HESS J1857+026 & 258.64 & 2.67 \\ 
\cline{1-4}
\end{tabular}
\caption{Sources in Stacking Catalogs, source type, name, Right Ascension ($\alpha$) and Declination ($\delta$) are listed.}
\label{table:source_list_stacking}
\end{table}

\clearpage
\section*{Full Author List: IceCube Collaboration}




\scriptsize
\noindent
R. Abbasi$^{17}$,
M. Ackermann$^{59}$,
J. Adams$^{18}$,
J. A. Aguilar$^{12}$,
M. Ahlers$^{22}$,
M. Ahrens$^{50}$,
C. Alispach$^{28}$,
A. A. Alves Jr.$^{31}$,
N. M. Amin$^{42}$,
R. An$^{14}$
% ...
\\

\noindent
% ...
$^{12}$ Universit{\'e} Libre de Bruxelles, Science Faculty CP230, B-1050 Brussels, Belgium \\
% ...
$^{14}$ Department of Physics and Laboratory for Particle Physics and Cosmology, Harvard University, Cambridge, MA 02138, USA \\
% ...
$^{17}$ Department of Physics, Loyola University Chicago, Chicago, IL 60660, USA \\
$^{18}$ Dept. of Physics and Astronomy, University of Canterbury, Private Bag 4800, Christchurch, New Zealand \\
% ...
$^{22}$ Niels Bohr Institute, University of Copenhagen, DK-2100 Copenhagen, Denmark \\
% ...
$^{28}$ D{\'e}partement de physique nucl{\'e}aire et corpusculaire, Universit{\'e} de Gen{\`e}ve, CH-1211 Gen{\`e}ve, Switzerland \\
% ...
$^{31}$ Karlsruhe Institute of Technology, Institute for Astroparticle Physics, D-76021 Karlsruhe, Germany  \\
% ...
$^{42}$ Bartol Research Institute and Dept. of Physics and Astronomy, University of Delaware, Newark, DE 19716, USA \\
% ...
$^{50}$ Oskar Klein Centre and Dept. of Physics, Stockholm University, SE-10691 Stockholm, Sweden \\
% ...
$^{59}$ DESY, D-15738 Zeuthen, Germany \\
% ...

\subsection*{Acknowledgements}

\noindent
USA {\textendash} U.S. National Science Foundation-Office of Polar Programs,
U.S. National Science Foundation-Physics Division,
U.S. National Science Foundation-EPSCoR,
Wisconsin Alumni Research Foundation,
Center for High Throughput Computing (CHTC) at the University of Wisconsin{\textendash}Madison,
Open Science Grid (OSG),
Extreme Science and Engineering Discovery Environment (XSEDE),
Frontera computing project at the Texas Advanced Computing Center,
U.S. Department of Energy-National Energy Research Scientific Computing Center,
Particle astrophysics research computing center at the University of Maryland,
Institute for Cyber-Enabled Research at Michigan State University,
and Astroparticle physics computational facility at Marquette University;
Belgium {\textendash} Funds for Scientific Research (FRS-FNRS and FWO),
FWO Odysseus and Big Science programmes,
and Belgian Federal Science Policy Office (Belspo);
Germany {\textendash} Bundesministerium f{\"u}r Bildung und Forschung (BMBF),
Deutsche Forschungsgemeinschaft (DFG),
Helmholtz Alliance for Astroparticle Physics (HAP),
Initiative and Networking Fund of the Helmholtz Association,
Deutsches Elektronen Synchrotron (DESY),
and High Performance Computing cluster of the RWTH Aachen;
Sweden {\textendash} Swedish Research Council,
Swedish Polar Research Secretariat,
Swedish National Infrastructure for Computing (SNIC),
and Knut and Alice Wallenberg Foundation;
Australia {\textendash} Australian Research Council;
Canada {\textendash} Natural Sciences and Engineering Research Council of Canada,
Calcul Qu{\'e}bec, Compute Ontario, Canada Foundation for Innovation, WestGrid, and Compute Canada;
Denmark {\textendash} Villum Fonden and Carlsberg Foundation;
New Zealand {\textendash} Marsden Fund;
Japan {\textendash} Japan Society for Promotion of Science (JSPS)
and Institute for Global Prominent Research (IGPR) of Chiba University;
Korea {\textendash} National Research Foundation of Korea (NRF);
Switzerland {\textendash} Swiss National Science Foundation (SNSF);
United Kingdom {\textendash} Department of Physics, University of Oxford.

\end{document}